# Modeling the microscopic electrical properties of

# thrombin binding aptamer (TBA)

# for label-free biosensors

Eleonora Alfinito[1], Lino Reggiani[2], Rosella Cataldo[2], Giorgio De Nunzio[2], Livia Giotta[3], Maria Rachele Guascito[3]

[1]*Dipartimento di Ingegneria dell`Innovazione. Università del Salento, via Monteroni, Lecce, Italy*
[2]*Dipartimento di Matematica e Fisica "Ennio de Giorgi", Università del Salento, via Monteroni, Lecce, Italy*
[3] *Dipartimento Di.S.Te.B.A. , Università del Salento, via Monteroni, Lecce, Italy*

## Abstract

Aptamers are chemically produced oligonucleotides, able to bind a variety of targets such as drugs, proteins and pathogens with high sensitivity and selectivity. Therefore, aptamers are largely employed for producing label-free biosensors, with significant applications in diagnostics and drug delivery. In particular, the anti-thrombin aptamers are biomolecules of high interest for clinical use, because of their ability to recognize and bind the thrombin enzyme. Among them, the DNA 15-mer thrombin-binding aptamer (TBA), has been widely explored concerning both its structure, which was resolved with different techniques, and its function, especially about the possibility of using it as the active part of biosensors. This paper proposes a microscopic model of the electrical properties of TBA and the aptamer-thrombin complex, combining information from both structure and function. The novelty consists in describing both the aptamer alone and the complex as an impedance network, thus going deeper inside the issues addressed in an emerging electronics branch known as proteotronics. The theoretical results are compared and validated with Electrochemical Impedance Spectroscopy measurements reported in the literature. Finally, the model suggests resistance measurements as a novel tool for testing aptamer-target affinity.

**I. Introduction**

Aptamers are target-specific single-stranded DNA, RNA or peptide sequences mainly generated on demand by an *in vitro* selection and amplification method called Systematic Evolution of Ligands by EXponential Enrichment (SELEX), which involves repetitive cycles of binding, recovery and amplification steps (Yüce et al., 2015). These techniques ensure the production of oligonucleotides specific for targeted disease therapies (Zhou and Rossi, 2014) and medical diagnosis (Jo and Ban, 2016). Recently, computational approaches have been proposed for *in silico* selection of aptamers with a higher target-binding ability, to be used for clinical diagnosis and medicine (Bini et al., 2011). Aptamers can be used under a broad range of experimental conditions as a result of their thermal and chemical stability, furthermore, they can be readily synthesized in large quantities at relatively low cost (Yüce et al., 2015). This constitutes a great advantage compared to antibody (Schlecht et al., 2006) production which is laborious and requires animal sacrifice. Therefore, academic and commercial interest in the field is growing, thus promising a substantial progress over the coming years (Yüce et al., 2015).

Particular attention is devoted to the DNA 15-mer aptamer which binds the exosite II of thrombin, TBA, whose efficacy in inhibiting platelet thrombus formation has been known for a long time (Li, et al. 1994). Traditional therapies against blood coagulation are based on the use of heparin and direct inhibitors like hirudin and bivalirudin (Dahlbäck, 2005). The side effects of these drugs are well known, the most relevant is bleeding (Ni et al., 2011, Barbone et al., 2012). TBA is under clinical trial and its activity and side effects can be controlled by matched antidotes (Rusconi et al., 2004).

Really the therapeutic aptamers tested to date have shown good safety margins between the pharmacologically effective dose and the toxicologically established no-adverse-effect levels (Bouchard et al., 2010). Moreover, due to the facility of production and amplification, TBA, as well as many other aptamers, has been recently used in the design of several biosensors, also known as aptasensors ( Iliuk et al., 2011;  Jo and Ban, 2016; Shen et al., 2016; Wang et al., 2016). Aptasensors are interesting tools offering convenient methodologies for detecting and measuring the levels of specific proteins in biological and environmental samples whose detection, identification and quantification can be very complex, expensive and time consuming (Yüce et al., 2015).

Among aptasensors, a significant role is played by label-free biosensors, designed with high detection sensitivity and selectivity (Strehlitz et al., 2008). In particular, electrochemical transduction of biosensors using aptamers as bioreceptors include methods like Electrochemical Impedance Spectroscopy (EIS), differential pulse voltammetry, alternating current voltammetry, square wave voltammetry, potentiometry or amperometry (Rodriguez et al.,2005; Strehlitz et al., 2008).

.

Xu et al. (2005) proposed an EIS method for aptamer modified array electrodes as a promising label-free detection for immunoglobulins E (IgE). They found lower background noise, decreased non specific adsorption, and larger differences in the impedance signals due to the small size and simple structure of the aptamers in comparison to the antibody (Xu et al., 2005). Cai et al. (2006) proposed an aptamer-based biosensing assay for label-free protein detection and quantification by measuring the change in electrochemical impedance upon protein–aptamer complex formation, monitoring the interfacial electron transfer resistance with electrochemical impedance spectroscopy (EIS). TBA showed high binding affinity and specificity to its target protein and the impedance detection assay was able to achieve a reliable and sensitive quantification of thrombin (Cai et al., 2006).

Although much work has been made in last 20 years a deep understanding of the mechanisms of TBA-thrombin recognition and capture is far to be reached.

This paper proposes a microscopic model of the electrical properties of the aptamer-thrombin complex, combining information from both structure and function of the biomolecule. Hereafter we use the term biomolecule for indicating both the aptamer and the aptamer-protein complex.

In particular we analyze the results of the experiment performed by Cai et al. (2006) that investigated the *in vitro* response of TBA samples to different thrombin concentrations.

Those experimental data are here described within the framework of *proteotronics*, i.e. a structure-function representation of the aptamer, able to connect its electrical response at thrombin growing concentration, considering both its structure modification and its internal energy variation (Alfinito et al., 2015). In particular, we argue that the single aptamer electrical response, as well as that of the single enzyme, are well reproduced by an impedance network analogue. The differences between these two networks are in the static electrical properties which are specific for the chosen biomolecule. Furthermore, the aptamer-enzyme complex is itself represented by an impedance network and the global electrical response is due to the cooperative action of the component networks.

In our opinion the proposed approach is more interesting than previous ones (Hou et al., 2006; Hou et al., 2007; Alfinito et al., 2013; Alfinito and Reggiani, 2014; Alfinito and Reggiani, 2015). It aims to verify whether a procedure born for proteins fits well also for different kinds of biomolecules, and whether a single impedance network is a valid model of the expected global electrical response of a biological compound.

This novel model takes advantage of recent data concerning some electrical properties of aptamers, i.e. polarizability (Šponer, et al., 2002) and resistivity (Ohshiro et al., 2012). The proposed electrical description gives a novel powerful tool for testing the aptamer-protein affinity. The results clearly show that a resistance measurement is able to recognize whether the protein-aptamer complex has been performed in the presence of $K^+$ or $Na^+$ ions. Since the complex formed in the presence of $K^+$ is more stable (Russo-Krauss et al., 2012), the resistance measurement seems to be a novel tool for resolving different affinities.

As far as we know, there is not another electrical microscopic modeling formulated for aptamers.

## 2.Materials

The thrombin binding DNA aptamer, 5'-GGT TGG TGT GGT TGG-3` is a SELEX product able to bind the blood coagulation enzyme thrombin. Its tertiary structure has been resolved with crystallographic (Schultze et al., 1994; Russo-Krauss et al., 2012) and computational (Bini et al., 2011) techniques.

In performing our investigations we considered:

*a.* TBA in the native state (without the enzyme, in the wild configuration), hereafter called TBA$_{nat}$;

*b.* TBA complexed (with the enzyme, in the binding configuration), hereafter called TBA$_{com}$;

*c.* TBA in the active state (without the enzyme, in the binding configuration), hereafter called TBA$_{act}$.

The 3D structure we used for describing TBA$_{nat}$ is the PDB entry 148D (Schultze et al., 1994; Berman et al., 2000), which is an NMR product with 12 different outputs. For TBA$_{com}$ we used the entries 4DII and 4DIH (Russo-Krauss et al., 2012; Berman et al., 2000), X-ray structures of the aptamer-protein complex resolved in the presence of potassium and sodium ions, respectively. Different effects exerted by $Na^+$ and $K^+$ on the inhibitory activity of TBA are also analyzed within our model. The $TBA_{act}$ structures were obtained by cutting off the protein from those of $TBA_{com}$.

The tertiary structures were used as the input of the model shown in Section 3.

### 3. Experiment and Theory

*Experiment*

The experiments performed by Cai et al. (2006), used Au electrodes functionalized with the aptamer 5'-thiol-TTT TTT GGT TGG TGT GGT TGG-3` whose core is TBA, and incubated with different concentrations of thrombin. After incubation electrochemical measurements were performed and the impedance spectra recorded. The plane representation of this spectrum is known as Nyquist plot. The semicircle diameter of the plot gives the electron transfer resistance $R_{ct.}$ The formation of the TBA-thrombin complex monotonically enlarges the value of $R_{ct.}$

Our aim is to reproduce these data by using the model proposed in the following.

### *Theory*

The electrical response is here described by using a model that conjugates structure and function of the single biomolecule. This choice aims to mimic a fundamental paradigm in biology, i.e. tertiary structure induces function and *vice versa*. Accordingly, the model resolution is at the level of the single amino acid/nucleobase. Previous experience gained with sensing proteins (olfactory receptors, opsins, enzymes) has shown that this level of resolution is sufficient for reproducing the electrical properties measured in macroscopic devices (Alfinito et al., 2015).

Biomolecule structure is reproduced by using a graph. By taking the $C_1$ and $C_\alpha$ carbon atoms as the centroids of each nucleobase/amino acid, their position is mapped into the position of a node of the graph. Then, two nodes are connected to form a link only if their distance is smaller than an assigned interaction radius, $R_C$.

Figure 1a represents the graph of the TBA$_{nat}$ (PDB entry 148D_1, $R_C$ =10.1 Å). Figure 1b shows the graph of the TBA$_{com}$ (PDB entry 4DII, $R_C$ =10.1 Å ).

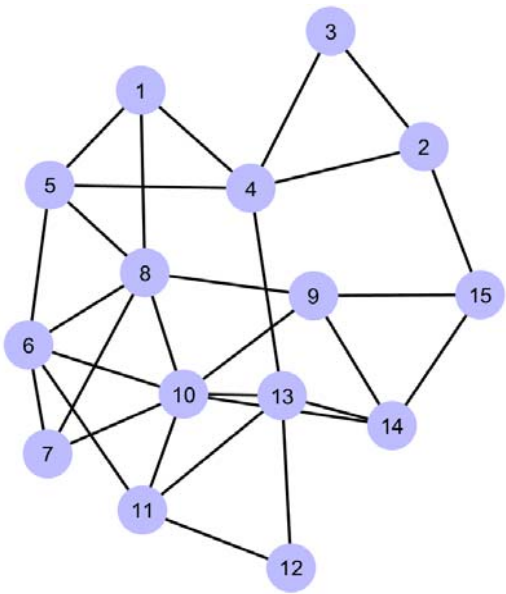
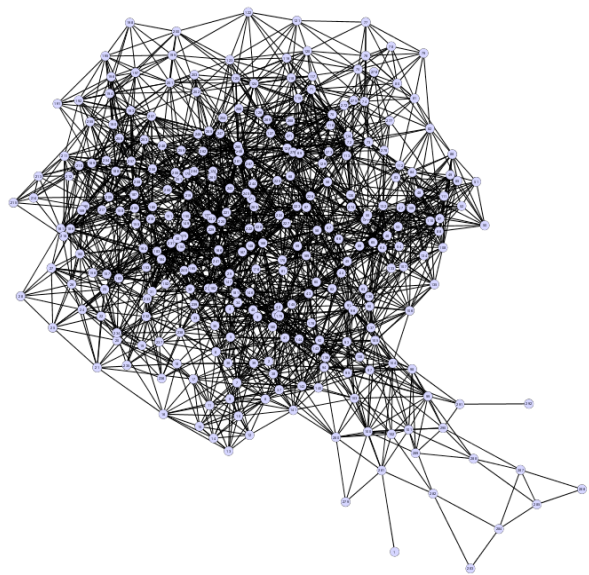

**Figure 1a. Graph of the TBA in the native state.** The 3D structure is taken by the PDB entry 148D_1, $R_C$=10.1 Å (Berman et al., 2000).

**Figure 1b. Graph of the TBA-thrombin complex.** The 3D structure is taken by the PDB entry 4DII, $R_C$=10.1 Å (Berman et al., 2000).

Each link mimics a specific electrical interaction, as observed in experiments (Hou et al., 2007; Alfinito et al., 2015). Specifically, the impedance measurements performed by Cai et al. (2006), are interpreted in terms of a macroscopic Randles cell made of 4 passive elements, i.e., the electron-transfer resistance, $R_{et}$, the solution resistance, the double-layer capacitance, $C_{dl}$, and the Warburg impedance, which is the linear part of the lowest frequency range and represents the diffusion-limited electron-transfer process (Cai et al., 2006).

In the actual experimental design the Warburg impedance is not foreseen, thus the thrombin injection changes the values of only two of these elements, say $R_{et}$ and $C_{dl}$, connected in parallel. Therefore, only $R_{et}$ and $C_{dl}$ were taken into account for modeling the electrical properties of the biomolecules. In other words, each link is replaced by an elemental impedance, describing the charge transfer and/or the charge polarization across the aptamer or the aptamer-protein complex. Since this is the relevant information for analyte detection, this is also the only electrical interaction we take into account. Therefore, each link is replaced by an elemental impedance, describing the charge transfer and/or the charge polarization across the aptamer or the aptamer-protein complex.

The elemental impedance is that of an RC parallel circuit:

$$Z_{a,b} = \frac{l_{a,b}}{A_{a,b}} \frac{1}{\rho_{a,b}^{-1} + i\varepsilon_{a,b}\varepsilon_0\omega} \qquad (1)$$

where $A_{a,b} = \pi(R_C^2 - l_{a,b}^2/4)$ is the cross-sectional area between spheres of radius $R_C$ centered on the $a$-th and $b$-th node, respectively; $l_{a,b}$ is the distance between these centers, $i = \sqrt{-1}$ is the imaginary unit, $\varepsilon_0$ is the vacuum permittivity, $\omega$ is the circular frequency of the applied voltage. The relative dielectric constant of the couple of $a$-th and $b$-th amino-acids, $\varepsilon_{a,b}$, is expressed in terms of the intrinsic polarizability of each isolated amino acid/nucleobase, $\alpha_{elec}$. In particular, $\varepsilon_{r\,i,j} = 1 + \frac{4\pi}{3}\left(\alpha_{elec,i} + \alpha_{elec,j}\right)/2$.

The resistivity of the $l$-th nucleobase is $\rho_l = \overline{\rho}\delta_l$, where $\overline{\rho}$ is the mean value calculated upon the AGCTU set (Ohshiro et al., 2012). The resistivity of the link between the $a$-th and the $b$-th nucleobase is defined as $\rho^N_{a,b} = \frac{\overline{\rho}(\delta_a + \delta_b)}{2}$. Analogous data are not given for amino acids, therefore we assume that their resistivities are proportional to $\overline{\rho}$, $\rho^A = \gamma\overline{\rho}$, with $\gamma$ the same for all the amino acids. The possible value of γ will be briefly discussed in the following. The link resistance of a couple of amino acids is, again, $\rho^A$. For both the relative dielectric constants and resistivities, we assume that previous formulas also hold in the aptamer-thrombin contact regions in which amino acids and nucleobases are connected between them. In particular, for the couple of the $a$-th nucleobase and $b$-th amino acid, the link resistance is:

$$\rho^{NA}{}_{a,b} = \frac{\overline{\rho}(\delta_a + \gamma\ )}{2}.$$

In other terms, we assume the concept of a linear superposition of the effects, i.e., each couple of nodes has a specific value of polarizability and resistivity which depends only on the selected nodes. The presence of the other nodes in the network or the position of the couple in the network does not affect this result. This also means that, in this study, the electric and dielectric effects of water and dissolved electrolytes are explicitly not considered. It is well known that this matter is quite hard to investigate also by using the effective methods of molecular dynamics calculations (Madan et al., 2016 ). In particular, the presence of solvent may play a role in the dielectric properties of the biomolecules and has been described by using both models which use distance-dependent dielectric constants and continuum electrostatic models (Gilson and Zhou, 2007) .The inclusion of solvent mainly affects the surface regions of the molecules and may be described like a non-uniform distribution of polarizabilities inside the biomolecule (Honig and Nicholis, 1995; Gilson and Zhou, 2007; Špooner et al., 2002). In particular, these values are larger close the surfaces (due to the presence of water) and smaller inside the protein. Finally, concerning the binding, the solvent effect is described by an increasing of the solvation energy of conformations with larger surface areas, i.e. by favoring the conformations with smaller surface areas (Gilson and Zhou, 2007).

For a biomolecule small as the TBA, probably the concept of surface area is not well defined and the choice of using the isolated polarizabilities for the whole aptamers is not in contrast with the presence of the solvent. Therefore, the same choice has been made for the protein and the aptamer-protein complex. On the other hand, the role of electrolytes in the formation of the aptamer- protein complex is here implicitly considered, by analyzing the electrical properties of complexes obtained in the presence of two different cations. Finally, as far as we know, an explicit role of solvent in the electrical transport properties is not verified.

The network is connected to an external bias by means of ideal contacts put on the first and last nucleobase and electrically solved by using standard techniques. In particular, by using the node Kirchhoff's law, the problem statement is expressed by a set of linear equations that are numerically solved by a standard computational procedure (Alfinito et al., 2008, 2010, 2011 ).

Concerning the process of TBA-thrombin conjugation, it is here described by mimicking protein activation due to the specific ligand. The mechanism of protein activation is a long time debated problem (Onuchic et al., 1997; Miyashita et al., 2005; Kobilka and Deupi, 2007; Park et al., 2008 ). The most recent models go beyond the *lock-and-key* scheme, by considering protein activation as the result of multiple pathways along and between energy funnels (Onuchic et al., 1997; Kobilka and Deupi, 2007; Alfinito and Reggiani, 2015). Each energy funnel is built up on a specific stable (*ground*) state, the native state (*native* funnel) or the binding state (*binding* funnel), for example. Analogously, we argue that when an aptamer folds from the molten state to the native state, its free energy, so as its conformational entropy, becomes smaller and smaller. This process is described by the aptamer sliding down an energy funnel (the *native* funnel), toward its minimum energy state (the *native* state). The interaction with thrombin may produce the binding or simply a variation of its free energy. In terms of the energy funnel description, the aptamer changes its state from native to binding, moving from the native funnel toward the binding funnel, or it stays in the initial funnel, but goes toward a higher energy state. In the impedance network analogue, both these kinds of dynamics go with a change of $R_C$. This is also in analogy with protein dynamics, since in going up and down among the states, the protein changes the number of its internal bonds, retaining only those useful for stabilizing its final configuration (Kobilka and Deupi, 2007).

Furthermore, when a *sample* receives its specific target, in general and if the target concentration is not huge, only a fraction of the aptamers binds the target and this is at the origin of the dose-response mechanism.

In particular, the dose-response given by experiments may be reproduced by using a Hill-like relation between the fraction of aptamer-protein complex, *f*, and the value of $R_C$ (Alfinito and Reggiani, 2015):

$$f = \frac{x^a}{b + x^a} \qquad\qquad (2)$$

where $x = (R_0 - R_C)/R_0$, $R_0$ is the value of $R_C$ corresponding to *f=0* i.e., a sample consisting only of TBAs in their native state, and *a* and *b* are numerical fitting parameters.

Finally, the sample resistance is given by:

$$r_{sample} = N\big[f \times r_{com} + (1-f) \times r_{nat}\big] \qquad\qquad (3)$$

where $N$ is the number of aptamers in the sample, $r_{nat}$ and $r_{comt}$ are the single aptamer and the single aptamer-thrombin complex resistances, respectively.

# 4. Results

## *a.        General results: single biomolecule*

As a first step, we calculated the global resistances of  TBA$_{nat}$ and TBA$_{act}$ . Numerical results are reported in Figures 2a, 2b.

We found that all the 12 native structures show a resistance smaller  than that of the active state, in particular for $R_C$ values close to 10 Å. Specifically, TBA$_{act}$, in the presence of both Na$^+$ and K$^+$,  exhibits a relative resistance variation, $rrv=r_{act}/r_{nat}$ -1 greater than 300%  when compared with the four native structures 148D_1, 148D_5, 148D_6, 148D_8. Such a greater resistance of TBA$_{act}$ with respect to TBA$_{nat}$ , indicates it has a more disconnected or dilated structure. Since we are interested in selecting one couple of structures with large resistance resolution, as shown by experiments, we focus on the first sequence of the native states, 148D_1, whose *rrv* value is intermediate among that of the four aforementioned native structures.  We notice that the results are quite independent on the kind of ions used for resolving the TBA$_{act}$ structures.

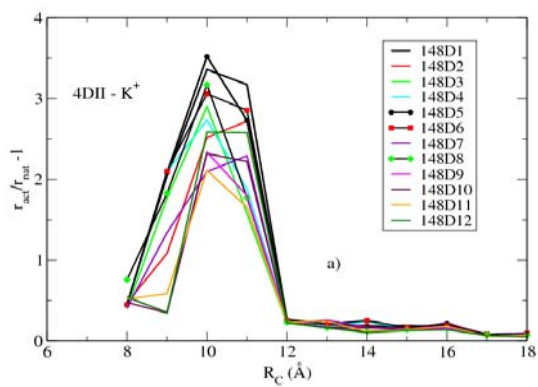

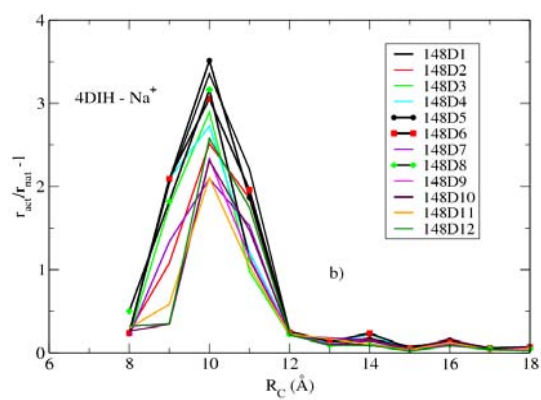

**Figure 2a. The relative resistances of the aptamer activated in the presence of potassium ions and in the native state.** The active state is taken by the PDB entry 4DII (Russo-Krauss et al., 2012), while the native states come from the PDB entry 148D (Berman et al., 2000). Calculations are performed at growing values of $R_C$. The structures of the native states which produce the largest differences with respect the activated state (1,5,6,8) are in black.

**Figure 2b. The relative resistances of the aptamer activated in the presence of sodium ions and in the native state.** The active state is taken by the PDB entry 4DIH (Russo-Krauss et al., 2012), while the native states come from the PDB entry 148D (Berman et al., 2000). Calculations are performed at growing values of $R_C$. The structures of the native states which produce the largest differences with respect the activated state (1,5,6,8) are in black.

Thrombin significantly changes these outcomes. To this purpose, Figures 3a and 3b report $rrv = r_{com}/r_{nat} - 1$ of the structures TBA$_{com}$ and 148D_1. The presence of the thrombin strongly reduces the differences observed in previous simulations, i.e., by using TBA$_{act}$ instead of

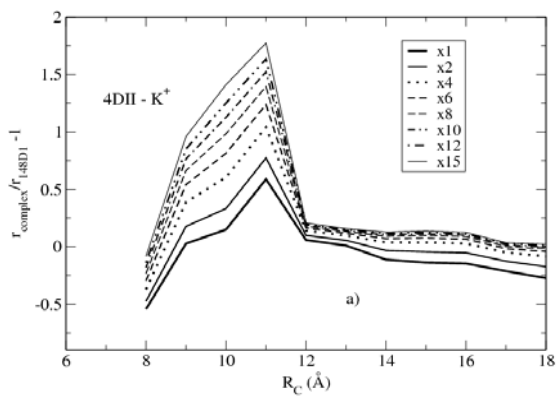

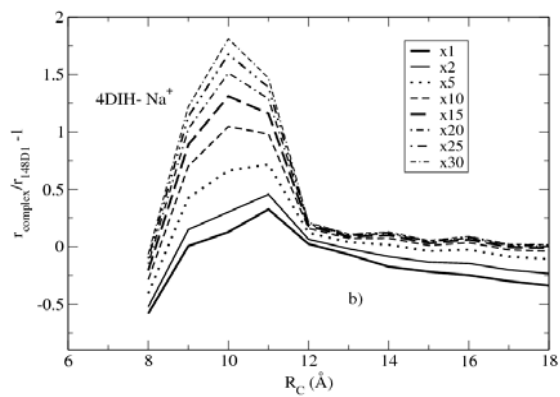

**Figure 3a. The relative resistances of the aptamer-thrombin complex in the presence of potassium ions and the native state.** The complex structure is taken by the PDB entry 4DII (Russo-Krauss et al., 2012), while the native structure comes from the PDB entry 148D_1 (Berman et al., 2000). Calculations were performed at growing values of $R_C$ and of $\gamma$, the proportionality factor of the amino acid resistivities (see text).

**Figure 3b. The relative resistances of the aptamer-thrombin complex in the presence of sodium ions and the native state.** The complex state is taken by the PDB entry 4DIH (Russo-Krauss et al., 2012), while the native state comes from the PDB entry 148D_1 (Berman et al., 2000). Calculations were performed at growing values of $R_C$ and of $\gamma$, the proportionality factor of the amino acid resistivities (see text).

TBA$_{com}$ , *rrv* falls from about 300% to 50% for the complex obtained in the presence of K$^+$ and to 25% for the complex obtained in the presence of Na$^+$.

As a matter of fact, thrombin completes the aptamer network, thus enhancing the number of connections and reducing the value of global resistance. This happens in different ways for the two structures, thus revealing that the binding is different. In particular, the formation of the aptamer-thrombin complex is promoted by the presence of K$^+$ (Russo-Krauss et al., 2012). Therefore, this result suggests resistance measurements as a complementary test of the aptamer-thrombin affinity.

We have also tested the role of the resistivity proportionality factor $\gamma$ , introduced in Section 3 (see Figures 3a and 3b). By increasing this value, the resistance of the aptamer-enzyme complex grows, as expected. On the other hand, this variation is not so massive to definitely suggest a value of $\gamma$ larger than 1.

*b.        Comparison with experimental dose response*

To compare theory and experiment, it is necessary to rescale the results obtained for the single biomolecule to the sample size. In particular, as described in the Theory subsection, we conjecture that, by increasing the thrombin concentration, the number of aptamers in the complex state becomes larger and that also the free energy of each aptamer becomes higher.

Accordingly, we select R$_0$=13.3 Å as the interaction radius useful to calculate the resistance of the single TBA in the native state. This value is the largest corresponding to the region in which, in agreement with experiments, the resistance of TBA$_{com}$ is larger than that of TBA$_{nat}$ (see Figures 3). The sample resistance is given by eq.(3) with *f*=0, up to $N$ . When the free energy of each aptamer changes, the R$_C$ value decreases (Alfinito et al., 2011) and its value is related to the value of *f* by

eq.(3). Finally, eq.(3) allows us to reproduce the resistance variation of the sample as given by experiments (dose-response). The rate $r/r_0$ of the sample $R_{ct}$ measured for different thrombin concentrations with respect the sample $R_{ct}$ without thrombin, the corresponding values of $R_C$ and the fraction , $f$, of aptamer-thrombin complexes are summarized in Table 1.

| Concentration (M) | $r/r_0$ | $R_C$ (Å) | $f$ |
|---|---|---|---|
| $1.0 \ 10^{-6}$ | 3.2 | 11.3 | 0.93 |
| $1.0 \ 10^{-8}$ | 2.1 | 11.5 | 0.90 |
| $1.0 \ 10^{-10}$ | 1.8 | 11.7 | 0.87 |
| $1.0 \ 10^{-12}$ | 1.2 | 12.7 | 0.26 |

**Table I. Summary of the experimental dose response results and theoretical data used to reproduce them.**

First column indicates the thrombin concentration in molar, by Cai et al. (2006) -experiment-; column 2 is the ratio between the resistance ($r_0$) of a sample not exposed to the thrombin, and the resistance ($r$) of a sample at the thrombin concentrations indicated in column 1 –experiment-. Column 3 gives the values of the interaction radius ($R_C$) used for calculating the fraction ($f$) of aptamer-protein complexes –theory-, column 4 gives the $f$ values (see eq.2) –theory-.

The corresponding Nyquist plots, as calculated within this model, are reported in Figure 4.

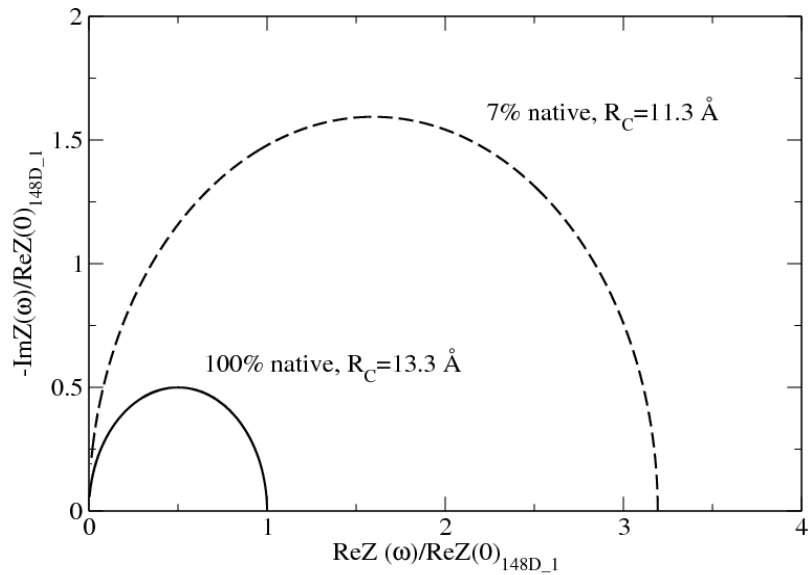

**Figure 4. The Nyquist plots of the aptamer-thrombin complex in the presence of potassium ions and the native state.** Dashed line represent the sample response given by 93% of TBA-thrombin complex (PDB entry 4DII), and 7% of TBA in the native (PDB entry 148D_1), the value of $R_C$ is 11.3 Å. Continuous line represents the response of a sample made of 100% of aptamers in the native state (PDB entry 148D_1), $R_C$ is 13.3 Å.

In particular, we report the Nyquist plot corresponding to a sample in which all the aptamers are in the native state, $R_C=R_0$=13.3 Å, and the Nyquist plot of a sample with the 93% of aptamers binding thrombin and $R_C$=11.3 Å, as given by eq.(2).

Finally, the plot of the Hill-like equation (2), is drawn in Figure 5. The fitting parameters are $a$=2.99 and $b$=2.7 $10^{-4}$.

In the inset of this figure we report the relative difference $r_T/r_R$ -$1$ of the sample resistance measured at different concentrations of thrombin, $r_T$, and the resistance of the reference sample (without thrombin), $r_R$, as function of the $R_C$ values we used for fitting the percentage of aptamer-thrombin complex.

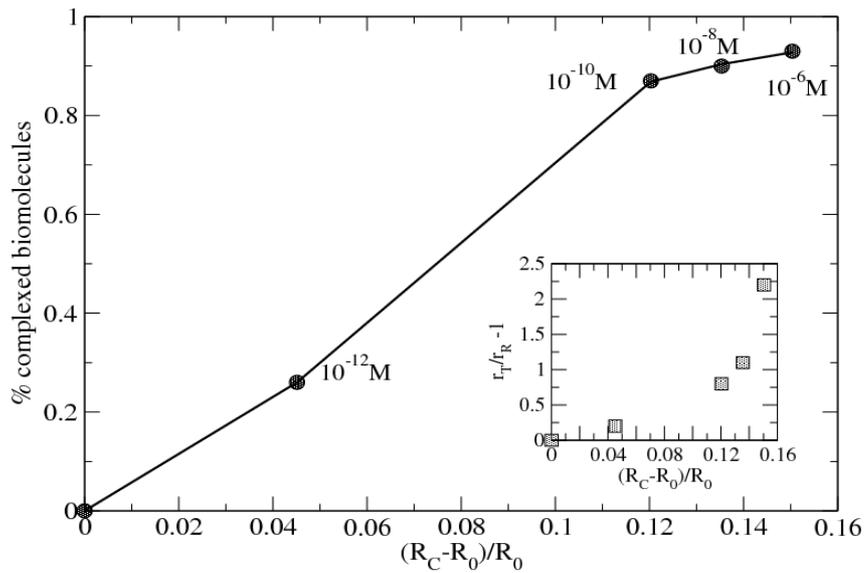

**Figure 5. The fraction of TBA-thrombin biomolecules used to fit the experiments (Cai et al., 2006).** Dots refer to the thrombin concentrations given in the experiments. Continuous line is the fit performed by using the Hill-like function described in the text. In the inset the relative resistance of the sample measured at different concentrations of thrombin, $r_T$, and that of the reference sample (without thrombin), $r_R$ as given by experiments, are reported in correspondence of the $R_C$ values we for the Hill fitting, as in the main figure.

## 4. Discussion and Conclusions

Aptamers are a new promising group of bioreceptors, because of their outstanding selectivity, sensitivity and stability (Strehlitz et al., 2008). Especially aptamer-based clinical label-free devices are widely recognized as the new frontier in diagnosis as well as in therapy. Unfortunately, poor information is at present available concerning the mechanisms of aptamer activation, so going deeper in this field is extremely challenging.

In this paper we detailed and tested a microscopic model useful for the description of the electrical properties of aptamers and aptamer-protein complexes, by using analogue impedance networks. The plug-in of aptamer and protein activates an electrical communication between them which results in the modification of the electrical response of the aptamer alone.

In particular, we analyzed the electrical response of three different structures, the aptamer in the native state, $TBA_{nat}$ , and the aptamer in the binding state, without the thrombin, $TBA_{act}$ , and with the thrombin, $TBA_{com}$. The relative resistance variation, $rrv = r_{act}/r_{nat} - 1$ calculated for $TBA_{nat}$ and $TBA_{act}$ is large, when compared with experiments but when the protein is added ($TBA_{act}$ is substituted by $TBA_{com}$) it becomes comparable with experiments. This result suggests that the resistance growth is not only a passivation effect due to the addition of a hydrophobic layer of proteins (Cai et al., 2006), but that the role of the specific structural docking is relevant, since the protein completes the impedance network and *reduces* the resistance which would have been measured in a sample with the aptamers activated but not bound to the protein.

Furthermore, the model predicts that a resistance measurement is a good test of the protein-aptamer affinity. The model is validated on EIS measurements performed by Cai et al. (2006) on TBA samples exposed to growing concentration of thrombin.

Finally, monitoring the change of $R_{et}$ obtained from EIS promises a sensitive and reliable protein label-free biosensor. The interpretation of these data by using a simple network impedance is a new success of proteotronics methods. All these findings could accelerate the development of label-free biosensors of practical relevance.

**Acknowledgements**

Prof. Filomena Sica, *Università di Napoli, Federico II*, is gratefully acknowledged for the useful discussions about the  aptamer structures.